\documentclass[12pt]{iopart}

\newcommand{\apj}{{\it ApJ}~}
\newcommand{\apjl}{{\it ApJL}~}
\newcommand{\mnras}{{\it MNRAS}~}

\newcommand{\eqb}{\begin{eqnarray}}
\newcommand{\eqe}{\end{eqnarray}}

\usepackage{iopams}  
\usepackage{bm}
\usepackage{graphicx}

\begin{document}

\title[Fermi acceleration at the laser produced shocks]{
Diffusive shock acceleration at laser driven shocks: 
studying cosmic-ray accelerators in the laboratory
}

\author{B Reville, A~R Bell and G Gregori}

\address{Clarendon Laboratory, Parks Rd., Oxford OX1 3PU, UK}
\ead{b.reville1@physics.ox.ac.uk}
\begin{abstract}

The non-thermal particle spectra responsible for the  
emission from many astrophysical systems are thought
to originate from shocks via a first order Fermi process 
otherwise known as diffusive
shock acceleration. The same mechanism is also widely believed to be responsible 
for the production of high energy cosmic rays. With the growing interest in 
collisionless shock physics in laser produced plasmas, the possibility of 
reproducing and detecting shock acceleration in controlled laboratory experiments
should be considered. The various experimental constraints that must be 
satisfied are reviewed. It is demonstrated that several currently operating
laser facilities may fulfil the necessary criteria to confirm the occurrence
of diffusive shock acceleration of electrons at laser produced shocks. 
Successful reproduction
of Fermi acceleration in the laboratory could open a range of possibilities,
providing insight into the complex plasma processes
that occur near astrophysical sources of cosmic rays.

\end{abstract}

\submitto{\NJP}
\maketitle

\section{Introduction}

The theory of diffusive shock acceleration was independently put
forward in four publications in the late seventies 
\cite{axfordetal77,bell78a,blandfordostriker78,krymskii77} as a 
mechanism to account for the non-thermal emission from astrophysical 
shocks and also as a possible explanation for the origin of cosmic rays.
A particularly attractive feature of this process is that, in the 
simplest test-particle theory, the accelerated particles naturally 
form power-law spectra consistent with those inferred from multi-wavelength
measurements. Despite the success of this theory in explaining observations
covering a broad range of astrophysical phenomena, from interplanetary and 
supernova remnant shocks to radio galaxy hotspots and $\gamma$-ray bursts, 
a complete understanding of the mechanism is still lacking.
Theoretical models and numerical simulations have developed rapidly over
the last 30 years, however, the limited data provided by in-situ 
satellite measurements at shocks in our solar system make verification 
of these models difficult. The ability to successfully perform experiments
to study diffusive shock acceleration in controlled 
laboratory environments would represent a major advance in the field.

Modern high-power laser facilities provide the means to generate strong 
shocks in the laboratory. These facilities are already being used to 
perform experiments in a parameter range where scaling relations may
be used to apply results to plasma physics studies of astrophysical relevance 
\cite{woolseyetal01,remingtonetal,gregorietal12}. The conditions required to make 
comparisons with astrophysical shocks has been the focus of several papers, 
e.g. \cite{ryutovetal99, drake00,woolseyetal01}. While several laser-plasma 
experiments have focussed on the generation and analysis of collisionless shocks 
\cite{courtois04,kuramitsu,Rossetal12,Kugland12}, to date there has 
been no report of a successful detection of shock acceleration. 
We review the relevant results of
the theory primarily in the context of laboratory laser driven shocks.
The extension to other plasma experiments, such as on the plasma railgun 
device at the Los Alamos National Laboratory \cite{Hsuetal_Hedla} is straightforward.
The constraints on plasma parameters
necessary to accelerate and more importantly, detect 
energetic particles are discussed in detail. It is demonstrated that the required 
conditions can in principle be satisfied for several current laser facilities, 
and that experiments designed to detect accelerated particles could be carried out 
in the very near future.

\section{Shock acceleration}
\label{sect:acc}

The acceleration of particles in a conducting fluid requires an electric field.
For laboratory plasmas, neglecting resistivity, a typical form for Ohm's law is
$\bm{E}=-\bm{u}_{\rm e}\times\bm{B}-\bm{\nabla}p_{\rm e}/n_{\rm e}e$,
with $u_{\rm e},~p_{\rm e}$ and $n_{\rm e}$ the electron fluid velocity, 
pressure and number density respectively.
On length scales $L \gg v_{\rm th,e}^2/u_0\omega_{\rm ce}$, 
where $v_{\rm th,e}$ and $\omega_{\rm ce}$ are the electron thermal 
velocity and cyclotron frequency respectively, and $u_0$ a characteristic velocity
of the background fluid, the pressure gradient can be neglected, and the
large scale electric field vanishes in the local plasma fluid frame.
This is the situation usually considered in astrophysical plasmas.
Since the fluid velocity is seldom
uniform on large scales, the electric field does not vanish globally, and, 
as a consequence, a
charged particle with sufficient momentum to decouple from 
the thermal plasma associated with the local fluid motion, 
can sample the global electric field
and accelerate to higher energy. This is the underlying principle of Fermi 
acceleration \cite{fermi}. While this process is typically quite slow
in the presence of small velocity fluctuations, at shock fronts,
where the incoming fluid is abruptly decelerated and compressed, 
the acceleration can be very rapid. The process by which a distribution of particles 
repeatedly samples this velocity jump is known as diffusive shock
acceleration (for detailed 
reviews, see e.g. \cite{blandfordeichler87, drury83}).

The efficacy of the acceleration hinges on a particle's ability to cross the 
shock surface many times, since the increase in energy on each crossing 
is relatively small. For a shock velocity $u_{\rm sh}$ and particle 
velocity $v~(\gg u_{\rm sh})$, the fractional increase in energy on each crossing is
of order $u_{\rm sh}/v$. 
Acceleration to high energies relies on a number of conditions. 
First, as mentioned above, 
the particle must have a sufficiently large initial momentum to
\emph{escape} from the thermal pool and overtake the shock.
This is the so-called injection problem, and remains a topic of ongoing 
investigation (see \cite{kirkdendy01, amanohoshino07, riquelme} and references therein). Second, the
scattering must be sufficiently frequent to maintain
near isotropic particle distributions. For acceleration to proceed
these scatterings must be mediated by quasi-elastic
interactions with magnetic fluctuations, as opposed to 
inelastic Coulomb collisions with other particles.
The scattering fluctuations in a laboratory setting can be produced 
for example via Weibel instability in the shock layer \cite{sagdeev66, kato08},
or hydromagnetic waves excited by shock reflected or accelerated particles
\cite{bell78b, bell04}. 
If a small fraction of the particles crossing the shock are heated to super-thermal
velocities, provided these particles are maintained approximately isotropic, 
the probability of a particle interacting with the shock multiple times is
high.

\subsection{Acceleration time}
\label{sect:acctime}

Assuming the above mentioned conditions can be satisfied, at an idealized 
shock, the resulting acceleration timescale is \cite{drury83} 
\eqb
\label{eq:tacc}
t_{\rm acc} = \frac{3}{u_{\rm u}-u_{\rm d}}
\left(\frac{D_{\rm u}}{u_{\rm u}}+\frac{D_{\rm d}}{u_{\rm d}}
\right)\enspace,
\eqe
where $u_{\rm u,d}$ are the upstream and downstream  flow velocities
as measured in the shock rest frame, and $D_{\rm u,d}$ the 
corresponding energy dependent diffusion coefficient in the direction normal
to the shock surface. The diffusion coefficient can vary quite significantly
depending on the inclination of the shock normal with respect to 
the local magnetic field. From scaling arguments,
it is clear that fast shocks with 
small diffusion coefficients are more rapid accelerators.
Since cross-field diffusion is much less effective than diffusion 
along magnetic field lines ($D_{\bot}\ll D_\|$),
perpendicular shocks, ie. shocks for which the incoming magnetic field 
is in a direction perpendicular to the direction of the shock's motion,
can have considerably shorter acceleration times \cite{jokipii87}.
From an experimental perspective, 
minimizing the acceleration time is crucial, since as we will
show in section \ref{sec:nrg}, the time window for making detections is narrow 
even in the most optimistic scenario. In addition, the acceleration rate is 
always competing with Coulomb collisions at low energies. 
We therefore focus on the acceleration of particles at 
perpendicular shocks.

In the quasi-linear approximation \cite{jokipii66}, the diffusion 
can be separated into two orthogonal components, parallel and perpendicular to 
the mean field:
\eqb
D_\|=\frac{D_{\rm B}}{\xi} \enspace \mbox{   and   } \enspace
D_\bot=\frac{\xi D_{\rm B}}{1+\xi^2} \nonumber
\eqe
where $\xi=(\omega_{\rm g}\tau_B)^{-1}
<1$ is the ratio of the effective collision rate, $\tau_B^{-1}$, 
to the gyrofrequency, and 
\eqb
D_{\rm B} = \frac{mcv^2}{3eB} 
\eqe
is the (non-relativistic) Bohm diffusion limit, corresponding to 
roughly one scattering per gyroperiod. Here $\tau_B$ is the mean time between 
collisions on magnetic fluctuations, not to be confused with Coulomb interactions.
The value of 
$\xi$ will depend on the value of the background field and 
also the level and scale of magnetic fluctuations. A typical scale for 
fluctuations in the plasma is the electron collisionless skindepth,
although magnetised shock experiments have shown evidence for structures
on the scale of the hot electrons' gyroradius \cite{courtois04}. 
For typical laboratory conditions, e.g. $B=1$T,
$n=10^{16}~{\rm cm}^{-3}$, these scales are comparable at electron 
temperatures of 400 eV (see section \ref{sec:inj}).

For perpendicular shocks, it is the details 
of the magnetic fluctuations in the immediate vicinity of 
the shock that determines the acceleration. If 
scattering is too weak ($\omega_{\rm g}\tau_B\gg1$), 
particles are tied to the field lines and are advected downstream
preventing further interaction with the shock, and thus are not accelerated 
efficiently \cite{belletal11}. 
If the scattering is too strong, ($\omega_{\rm g}\tau_B \rightarrow1$), the 
diffusion is approximately isotropic. In this case the direction of the 
magnetic field becomes insignificant, particles can make long excursions 
both upstream and downstream before returning to the shock, and the acceleration time 
correspondingly increases.
The optimal value for acceleration at a perpendicular shock is 
$\omega_{\rm g}\tau_B = v/u_{\rm sh}$ \cite{jokipii87} although in
practice it may be less than this. We will adopt the optimal value for most 
of the calculations that follow, since at high energies, provided 
$\omega_{\rm g}\tau_B\gg 1$, geometry plays the limiting role, while at 
low energies $v/u_{\rm sh}$ is not very large and any correction is likely to be
of order unity.

\subsection{Maximum energy}

In the absence of radiative (synchrotron, Bremsstrahlung
 etc.) or adiabatic cooling, the maximum energy to which 
a particle can be accelerated is limited either by time or geometry. For  
acceleration at perpendicular shocks, we can demonstrate that the acceleration
time is shorter than the hydrodynamic time, and thus
the geometry plays the key role.
In astrophysics, this limit is commonly referred as the Hillas 
criterion \cite{hillas},
and corresponds to the maximum potential difference a particle can experience in
a system of fixed size 
\eqb
\label{eq:maxnrg}
T_{\rm max} \approx eER_{\rm sh} \approx e|\bm{B}|(u_{\rm sh}/c) R_{\rm sh}
\approx B_4 ~u_{\rm sh,7}~ R_{\rm sh} ~{\rm keV}
\eqe
where $B=B_4 \times 10^4$~G is the strength of the magnetic field, 
$u_{\rm sh}=u_{\rm sh,7} \times 10^7~{\rm cm~s}^{-1}$ is the shock velocity
and $R_{\rm sh}$ is the shock radius in cm. The same subscript notation is
used throughout the paper. For perpendicular shocks, inserting 
(\ref{eq:maxnrg}) into (\ref{eq:tacc}) gives 
$u_{\rm sh,7}~t_{\rm acc}(T_{\rm max}) \approx \xi R_{\rm sh}$,
such that our assumption on geometry limited acceleration
is consistent (since $\xi < 1$).
To make an unambiguous
detection of shock accelerated particles, we require as large a separation 
as possible between $T_{\rm max}$ and the injection energy.
This is discussed further in the next section.  

\section{Practical considerations}

Most astrophysical shocks of interest are highly collisionless systems.
Such shocks are generally believed to be excellent particle accelerators.
Achieving similar collisionless conditions 
in the laboratory however is not straightforward, and places 
stringent limitations on 
experiments designed to reproduce shock acceleration. While the total 
energy will depend on the maximum laser energy that a given system can provide,
the external conditions must also satisfy a number of criteria.
As outlined in the previous section, a mean field is necessary to prevent 
particles escaping far upstream. There have been a number of experiments
designed to study collisionless magnetised shocks \cite{woolseyetal01,courtois04},
using conditions relevant for scaling to supernova shocks. An interesting 
outcome of these experiments of vital importance here was the observation
that very little penetration of the magnetic 
field into the shocked plasma occurred. Compression of the magnetic field 
is almost certainly required to accelerate rapidly to high energies 
(see Fig \ref{fig:1}).

Since the focus here is not to produce scaled versions of supernovae, but rather
to study acceleration from first principles, and to help the magnetic field lines
penetrate the plasma, we consider an alternative 
experimental set-up to that of \cite{woolseyetal01,courtois04}
based on the previous experimental design of
\cite{gregorietal12}. In this design, the laser or lasers are focused onto a 
central target in a neutral-gas filled chamber, driving a quasi-spherical shock
into the ambient gas. Since the magnetic 
field already penetrates the ambient gas before the shock arrives, provided the
gas is ionised sufficiently far upstream of the shock, there should be no
issue with magnetic field penetrating the plasma or prevented from
being advected downstream.  

We now discuss in detail the constraints on density and magnetic field
strength for a given laser energy in the context of shock acceleration.

\subsection{Energy budget}
\label{sec:nrg}

The total energy available to the expanding shock is the principal limiting
factor for any shock acceleration experiment, as already evident 
from equation (\ref{eq:maxnrg}). For a spherical explosion, the
shock undergoes a short-lived ballistic expansion before evolving to
a Sedov-Taylor expansion $R_{\rm sh}\propto t^{2/5}$. In this phase, the
shock is already decelerating $u_{\rm sh} \propto t^{-3/5} \propto R_{\rm sh}^{-3/2}$.
Inserting this scaling into equation (\ref{eq:maxnrg}), for a constant 
magnetic field, the maximum energy decreases with time 
$\propto t^{-1/5}$. Thus, once the expansion velocity is determined,
the maximum energy at a given distance can be inferred 
from equation (\ref{eq:maxnrg}).

For a given experiment, the total deposited laser energy is divided up into the
production of radiation, ionization of the external medium, magnetic energy, 
thermal energy and shock accelerated particles, if present. 
For low Z gases such as Helium at densities $n\ll 10^{18} {\rm~cm}^{-3}$,
the radiative cooling is dominated by Bremsstrahlung, however, for
the experimental set-up we consider here, the cooling 
time is much longer than the dynamical time \cite{ryutovetal99}.
At low densities, the ionization potential is typically smaller than the 
thermal energy density, and so can be neglected.
Hence, to a good approximation we can assume a typical Sedov-Taylor 
solution
\eqb
\label{eq:totalnrg}
R_{\rm sh} =  C_0 \left(\frac{E_0t^2}{\rho_{\rm ext}}
\right)^{1/5}
\eqe
where $E_{0}$ is the total energy in the blast-wave,
$\rho_{\rm ext}$ the gas density in the target chamber
and for a Helium gas, the numerical constant is $C_0\approx1.15$.
For laser plasma interactions, while a large fraction of the laser energy
is absorbed by the target, the fraction of this energy
that goes into the blast-wave is uncertain. 
Comparing with similar spherical blast-wave experiments 
such as those carried out in \cite{gregorietal12}, $E_0 = 0.01E_{\rm laser}$ 
provides an excellent fit to the data, however, careful target design may increase
this number. We parametrise the fraction of the laser energy
transferred to the blast-wave as $E_0=10^{-2}\eta_{-2}E_{\rm laser}$ and
leave $\eta_{-2}$ as a free parameter. 
The shock velocity at distance $R_{\rm sh}$ is therefore
\eqb
\label{eq:shockvelocity}
u_{\rm sh,7} \approx 6\frac{\eta_{-2}^{1/2}
E_{\rm kJ}^{1/2}}{n_{16}^{1/2}R_{\rm sh}^{3/2}} 
\eqe
where $E_{\rm kJ}$ is the total laser energy in kilo Joules 
and $n_{16}$ the external gas number
density in units $10^{16}~{\rm cm}^{-3}$.
Combining with equation (\ref{eq:maxnrg}), and assuming a Helium filled 
target chamber, the maximum energy as a function of distance is
\eqb 
\label{eq:Emax}
T_{\rm max} \approx 6~\eta_{-2}^{1/2}~B_4 
~E_{\rm kJ}^{1/2}~n_{16}^{-1/2}~R_{\rm sh}^{-1/2} ~{\rm keV}
\eqe

As pointed out by Drake (2000) \cite{drake00}, a slowly diverging 
plasma expansion such as the \lq\lq laser-driven rocket\rq\rq, can drive
a shock at high velocity over a longer distance. However, we note that the 
gain in time is nearly balanced by the reduction in $T_{\rm max}$, 
since the maximum energy is determined by the lateral extent of the shock. 
Alternatively a hemispherical blast-wave could be generated,
although the maximum energy would only change by a factor of $\sim\sqrt{2}$.  

\subsection{Magnetic field}

Following the discussion in section \ref{sect:acctime}, it is clear that
having a strong magnetic field is advantageous. It increases both the 
acceleration rate and the maximum energy. However, even in the presence of
efficient scattering, there is a limit on the 
maximum magnetic field that permits acceleration.
This additional constraint on the external medium, is that the magnetic pressure 
$B^2/4\pi$, should not exceed the shock ram pressure $\rho_{\rm ext} u_{\rm sh}^2$.
This is equivalent to saying that the shock is super-Alfv\'enic, $M_A>1$.
Inserting numerical quantities, the necessary condition is
\eqb
\label{eq:mach1}
n_{16}^{1/2}~u_{\rm sh,7}~B_4^{-1} > 1\enspace.
\eqe

Since the expected density and magnetic field are approximately 
constant in the experiment, this condition will ultimately be violated 
when the shock has decelerated appreciably.

\subsection{Collisions and magnetic diffusivity}
\label{sec:magdiff}

A pre-requisite for shock acceleration is that the particles approximately 
conserve energy between shock crossings. For this to occur, Coulomb collisions 
must be negligible for the accelerating particles. For rapid acceleration
at perpendicular shocks the pathlength on either side of the shock is 
on the order of its gyroradius. 
The ratio of the Coulomb mean free path, $\lambda_{\rm mfp}$, 
of an electron to its gyroradius 
is, assuming a Coulomb logarithm $\ln\Lambda\sim10$ \cite{NRL},
\eqb
\label{eq:collisions}
\frac{\lambda_{\rm mfp}}{r_g}\approx 0.6~B_4~n_{16}^{-1}~T_{\rm e}^{3/2}
\eqe
where
$T_{\rm e}$ is the electron temperature in eV. 
Since this ratio grows rapidly with electron energy, it is sufficient 
to demonstrate that collisions are negligible at the injection energy.
The equivalent ratio for protons is approximately $\sqrt{m_{\rm e}/m_{\rm p}}$
times smaller, making the acceleration of protons far more difficult.
To achieve acceleration, we require the associated acceleration time at
the injection energy
be shorter than the inverse of the Coulomb scattering frequency. 
Adopting the previously discussed optimal scattering rate 
$\xi=u_{\rm sh}/v$, the necessary condition for Coulomb scattering to be 
unimportant can be expressed as $\lambda_{\rm mfp}/r_{\rm g}\gg v/u_{\rm sh}$, 
which together with equation (\ref{eq:collisions}) gives \footnote{ 
Repeating the same calculation for protons, we require 
$B_4~n_{16}^{-1}~T_{\rm p} u_{\rm sh,7} \gg 100$. While this condition can 
be satisfied at a fast shock, any accelerated protons 
will be indistinguishable from simply shock heated protons.} 
\eqb
\label{eq:collision1}
B_4~n_{16}^{-1}~T_{\rm e}~u_{\rm sh,7} \gg 0.1 \enspace.
\eqe

If the scattering rate is closer to unity, the acceleration rate 
decreases, and the minimum injection energy must increase accordingly. In the other 
extreme, where cross field diffusion becomes negligible ($\xi\ll u_{\rm sh}/v$),
acceleration can not occur. In this scenario, seeding of electric and 
magnetic fluctuations may be achieved using fast electrons (see next section).

For a strong shock, the downstream temperature of the shocked ions,
in this case Helium, is according to the Rankine Hugoniot relations
$T_{\rm i} \approx 75 u_{\rm sh, 7}^2$ eV. The downstream 
electron temperature may be as much as $m_{\rm e}/m_{\rm p}$
times smaller than this, although there is considerable observational 
evidence that in collisionless shocks, this ratio ($T_e/T_i$) is closer to $0.1$
\cite{schwartzetal, vanadelsberg}. In addition, 
it seems likely that even for mildly collisional shocks, a 
small fraction of the electrons can still be heated to considerably higher 
energies due to the collective processes in the shock layer, and satisfying 
(\ref{eq:collision1}).

The role of collisions and the resulting magnetic diffusivity 
may also be important.
If the magnetic Reynolds number is not appreciably greater than unity, the
ability of the flow to distort and compress the magnetic field is severely 
limited. This can dramatically reduce the efficiency of shock acceleration.
Examination of (\ref{eq:collisions}) reveals however, that even modest temperatures 
of a few eV are sufficient to magnetise the electrons. Electron temperatures
similar and considerably larger than this have already been produced 
in previous experiments \cite{gregorietal12, parketal12}, and this 
is unlikely to present a serious limitation.

\subsection{Particle injection}
\label{sec:inj}

In both astrophysical and laboratory shocks,  
perhaps the biggest uncertainty in the theory of shock acceleration is 
how and in what quantity particles are lifted from the thermal background
and injected into the acceleration process \cite{kirkdendy01}. 
If the shock is indeed collisionless, we can put some estimates on 
critical length and energy scales of the problem.
For collisionless shock experiments, a crucial parameter is the 
collisionless skin-depth 
\eqb
\lambda_{\rm SD} = \frac{c}{\omega_{\rm pe}} \approx 
50~n_{16}^{-1/2}~\mu{\rm m}\enspace .
\eqe
With the gyroradius of an electron in such an experiment 
$r_{\rm g}\approx 75 T_{\rm keV}^{1/2}/B_4~\mu$m, it follows that electrons with
energy $T_{\rm e}\sim400 B_4/n_{16}$eV will interact resonantly with structures on
this scale, and it might be expected for a collisionless shock,
a fraction of particles will be naturally heated to such temperatures. 
For perpendicular shocks, there are 
a number of collective mechanisms believed to pre-heat the electrons e.g. Lower Hybrid 
waves \cite{bingham, McClementsetal}, whistler waves \cite{amanohoshino07,
riquelme}, or indeed the collective processes mediating 
the shock itself \cite{sagdeev66}. 
Recent kinetic simulations are advancing our knowledge of
different electron injection mechanisms at collisionless shocks \cite{amanohoshino07,
riquelme, gargate}, however, a complete theory is still lacking.
Nevertheless, there is a great deal
of observational evidence supporting the fact that the bulk electron temperature 
downstream can be on the order of $10\%$ that of the ions, which may be 
sufficient for electrons to escape upstream. In practice we only require 
a small fraction of the incoming electrons to achieve such high energies.

The uncertainty associated with injection is evident, and can not be relied 
upon. This clearly
emphasises the need to have an alternative mechanism in place should it be
needed. One such possibility is to inject a population of 
energetic electrons at early times. While this may be experimentally challenging
and entails a certain amount of fine tuning, it should be possible.
The generation of fast electrons using an external source is easily achieved by
irradiating an additional target. The mean kinetic energy and total flux of 
fast electrons are also straightforward to calibrate \cite{gitomer86,beg97}.
However, fast electrons are known to produce their own
electric and magnetic fields as they propagate, potentially  
influencing the acceleration of particles. Provided the fields 
produced saturate with total energy density less than that of the ambient 
magnetic field, the acceleration will still be dominated by the zeroth order 
field. This effect can be calibrated with control shots in the absence 
of a shock. However, the enhanced turbulence level may also increase the cross-field 
diffusion, thus reducing the acceleration rate. The magnitude of this effect 
is entirely model specific, but is not entirely unjustified in an astrophysical 
context. Most astrophysical shocks are known to excite instabilities upstream of 
the shock due to cosmic-ray or shock reflected ion currents.
While this is an interesting and topical area in cosmic-ray acceleration research, 
if the aim is to study the acceleration mechanism in its simplest test particle form,
the fast electron flux should be calibrated, so as to minimise this
effect. The fast electrons should also be timed to intersect 
with the shock as early as possible, subject to the condition $r_{\rm g}<R_{\rm sh}$.

\section{Necessary conditions for detection}

The necessary requirements to achieve shock acceleration have been detailed 
in the previous section. However, as has been regularly emphasised, the key 
objective is to produce an unambiguous detection. The main requirements are 
contained in Equations (\ref{eq:Emax}), (\ref{eq:mach1}) and (\ref{eq:collision1}).
Supplementing these conditions with the assumption that some fraction of the electrons
are heated to an energy $T_{\rm inj}$ approximately $10\%$ that of the shocked ions,
$T_{\rm inj} = 0.1\alpha T_{\rm i} = 7.5 \alpha u_{\rm sh,7}^2$eV
and shock velocity
$$u_{\rm sh,7} = 6\frac{\eta_{-2}^{1/2}E_{\rm kJ}^{1/2}}
{n_{16}^{1/2}R_{\rm sh}^{3/2}}~,$$
the velocity can be eliminated from the above conditions, and
the most relevant dimensionless quantities are:
$$
M_{\rm A} = 6\frac{\eta_{-2}^{1/2}E_{\rm kJ}^{1/2}}{B_4R_{\rm sh}^{3/2}} \enspace; \enspace
 \left.\frac{t_{\rm coll}}{t_{\rm acc}}\right|_{T_{\rm inj}}
 = 245 \alpha \frac{
B_4 \eta_{-2}^{3/2}E_{\rm kJ}^{3/2}}{ n_{16}^{5/2} R_{\rm sh}^{9/2}} 
\enspace; \enspace \frac{T_{\rm max}}{T_{\rm inj}} = 20
\frac{B_4 n_{16}^{1/2} R_{\rm sh}^{5/2}}{\alpha 
\eta_{-2}^{1/2} E_{\rm kJ}^{1/2}}\enspace .
$$

Taking reasonably conservative minimal requirements,
the experimental parameters must satisfy the following inequalities:

\begin{enumerate}
 \item Super-Alfv\'enic ($M_{\rm A} > 4$)
$$R_{\rm sh}<1.3\, \frac{\eta_{-2}^{1/3}E_{\rm kJ}^{1/3}}{B_4^{2/3}}\enspace{\rm cm}$$
 \item Injected electrons are collisionless ($t_{\rm coll} >10 t_{\rm acc}$)
$$R_{\rm sh}< 2\, \frac{\alpha^{2/9}
\eta_{-2}^{1/3}E_{\rm kJ}^{1/3}B_4^{2/9}}{n_{16}^{5/9}}
\enspace{\rm cm}$$
 \item significant gain ($T_{\rm max} > 10 T_{\rm inj}$)
$$ R_{\rm sh} > 0.75\, \frac{\alpha^{2/5}\eta_{-2}^{1/5}
E_{\rm kJ}^{1/5}}{B_4^{2/5}n_{16}^{1/5}} 
\enspace{\rm cm}$$
\end{enumerate}

As can clearly be seen, detecting electron acceleration using a kilo-Joule
facility has a small window for success, since a clear distinction between 
background and thermal particles sets a lower bound on the shock position, 
while the remaining conditions set upper bounds. As an example, 
consider an experiment where we achieve the conditions such that 
all normalised quantities ($E_{\rm kJ}$,
$B_4$ etc.) are unity. The above inequalities
are satisfied for $0.75~{\rm cm}<R_{\rm sh}<1.3~{\rm cm}$. Increasing the laser energy to values
relevant for facilities such as Omega, and in particular the 
National Ignition Facility, increases the available time
window, although the weak dependence of the upper and lower bounds on laser energy 
(to the $1/3$ and $1/5$ power respectively) imply only a marginal increase.

\section{Diagnostics}

While the necessary conditions for detection have been outlined, 
a method for measuring the presence of accelerated particles has 
so far not been discussed. 
To this end, accurate diagnostics of the plasma parameters are 
essential. The plasma density, temperature and 
magnetic field strength can be probed using standard techniques such 
as interferometry, Thomson scattering and Faraday rotation. In addition, 
Schlieren imaging can be used to determine the shock velocities.

\begin{figure}
\begin{center}

\includegraphics[height=0.35\textheight]{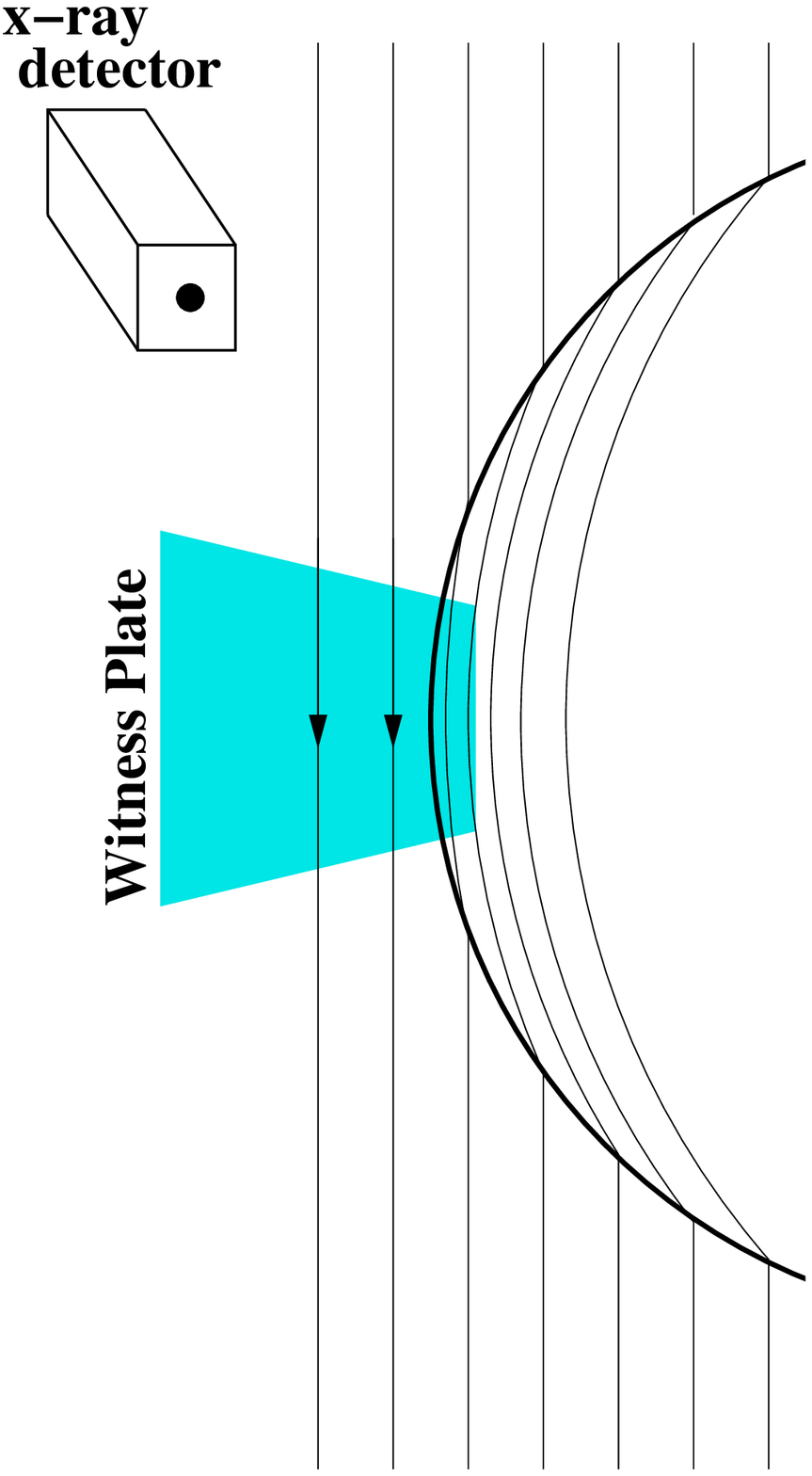} 
\includegraphics[height=0.33\textheight]{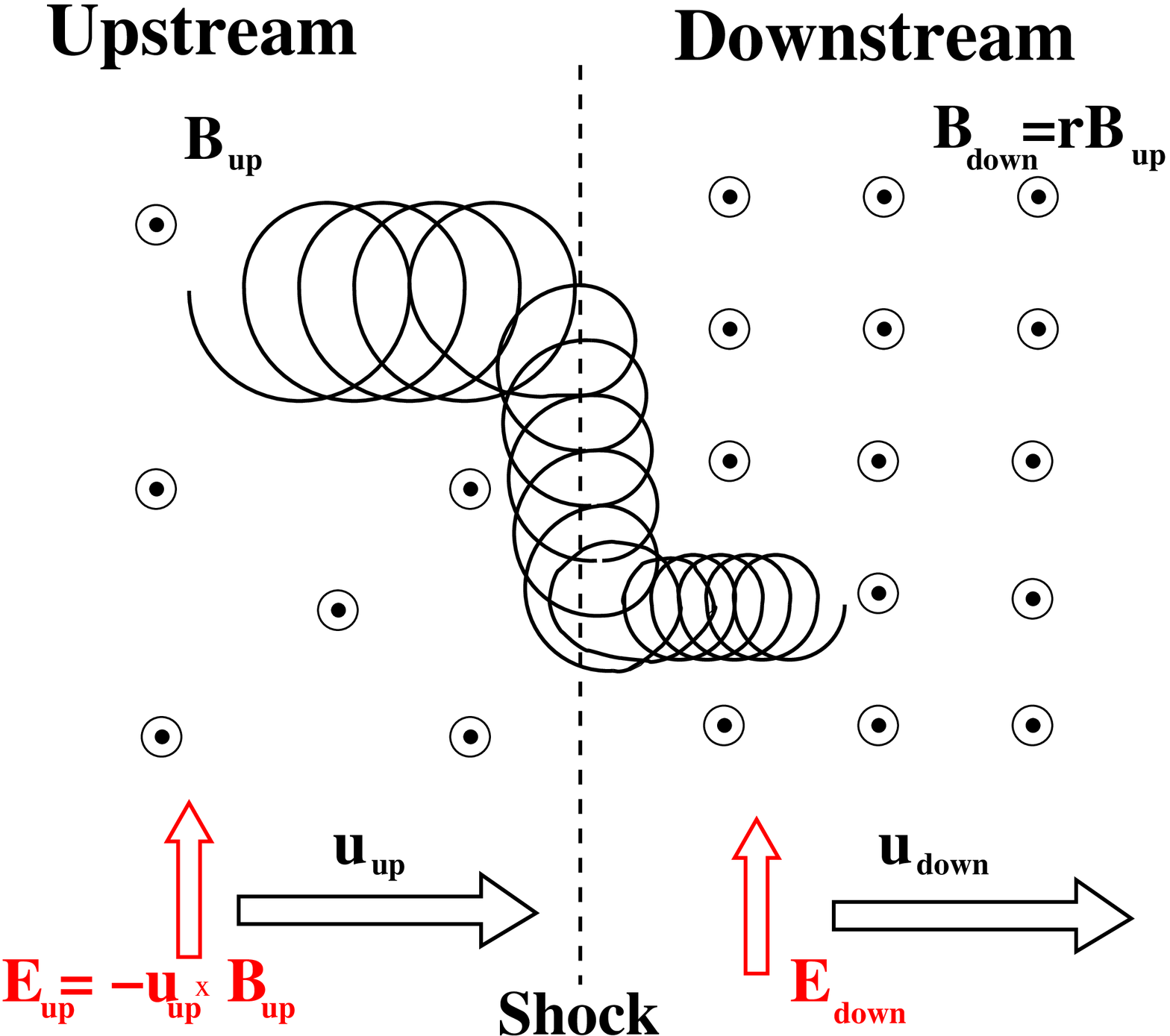} 

\end{center}
\caption{Left: Edge on view, showing spherical shock wave in uniform 
magnetic field. In the absence of magnetic diffusion, 
the tangential component of the magnetic field is compressed at a strong 
shock by a factor $r=(\gamma+1)/(\gamma-1)$. \\
Right: side view of a small section of the shock where field is purely
perpendicular. Electrons 
experience a grad B drift along the surface of the shock as a result 
of the magnetic field compression. In this example we have assumed no 
scattering, and the particles are accelerated by the shock drift mechanism
\cite{Webbetal}.
Small angle scattering on magnetic fluctuations allow some particles 
to \lq hug\rq~the shock surface for a hydrodynamic timescale (see section
\ref{sect:acctime}). The preferred direction of 
the accelerating electrons can be used to detect the presence of 
shock acceleration.}
\label{fig:1}
\end{figure}

Determining the presence of the non-thermal particles is more challenging.
The electrons should be detected in the range of $R_{\rm sh}$ determined
in the previous section, since at large radii the adiabatic losses of the 
particles in the expanding plasma is important.
Taking conditions (ii) and (iii) in the previous section,
the maximum energy electrons are expected in the range 
$$ 3.8 \frac{\eta_{-2}^{1/3}E_{\rm kJ}^{1/3}B_4^{8/9}}{n_{16}^{2/9}} 
<T_{\rm max}({\rm keV})
< 7~ \frac{\eta_{-2}^{2/5}E_{\rm kJ}^{2/5}B_4^{6/5}}
{\alpha^{1/5}n_{16}^{2/5}} $$
with the shock heated electrons $T_{\rm inj}$, also by condition (iii), an order of 
magnitude smaller. The maximum energy electrons are expected to fall in the range 
where it is possible to make use of the radiative Auger effect \cite{auger}. 
A high Z metal witness plate placed at
the appropriate location can thus be used as a probe of the accelerated 
electrons, as shown in left image of Figure \ref{fig:1}. Alternatively, the
target chamber plasma could be doped with high Z gas such as argon (with $n_{Ar} / n_{16} = f \sim 0.01$)  
that could show enhancement of inner shell emission near the shock. Both these
approaches may be quite sensitive to injection efficiency. Assuming an injection 
efficiency of $\chi\sim10^{-4}$, ie. $10^{-4}$ of the upstream electrons crossing 
the shock per unit time are injected into the acceleration process, and 
assuming the test particle solution for diffusive shock accelerated particles:
$dN(T)\propto T^{-2}dT$, the number of electrons at $T_{\rm max}$ is  
$$N(T_{\rm max}) \approx 10^{12}\chi_{-4}n_{16}\left(\frac{T_{\rm max}}{T_{\rm inj}}
\right)^{-2} \sim 10^{10}\chi_{-4}n_{16}
$$
where we have again taken $T_{\rm max}=10T_{\rm inj}$.
The Ar inner shell ionization cross section peaks at $\sim$5 keV ($\sigma_{iz} = 3\times10^{-21}$
cm${}^{2}$ \cite{SCHNEIDER1992}), and assuming a fluorescence yield $Y_K \sim 0.14$ \cite{watanabe1962},
the estimated number of photons collected within the detector solid angle $\Omega \sim 0.1$ sr is
$$ 
N_{ph} \approx 3\times 10^{25} f \, Y_K \, \chi_{-4} \, \Omega \, \sigma_{iz} \, R_{\rm sh}^4 \, n^2_{16},
$$
which gives $N_{ph} \sim 4$ -- $35$ for shock radii considered here.
While small, this number is sufficient for single shot detection, especially 
if a high throughput crystal monochromator
is used to enhance the signal to noise ratio.

Another possible technique that may be used to make a 
detection, involves taking advantage of the perpendicular 
geometry. At a perpendicular shock,
the acceleration has a directional bias, determined by the grad B drift,
as shown in figure \ref{fig:1}. The resulting asymmetry in the 
x-ray luminosity on opposite sides of the witness plate could be detected 
using two pinhole x-ray cameras. 

In all cases, care should be taken not to confuse any signal from 
fast electrons produced either from the laser-target interaction, or 
stray electrons from the external injection source. Again, control 
experiments can be used to confirm or invalidate any successful detections.

\section{Discussion}

The theory of diffusive shock acceleration has been incredibly successful in 
accurately modelling non-thermal radiation from astrophysical sources for several 
decades. While the evidence for its occurrence 
at numerous shocks in both astrophysical and space environments is convincing, there 
are many aspects that can not be fully understood from distant observations
and satellite measurements. In particular questions about injection, 
self-regulation, maximum energy, non-linear feedback and magnetic field amplification
are active areas of research in the theoretical community. The study of
collisionless shocks using high power lasers is a growing field
\cite{woolseyetal01,courtois04,kuramitsu,Rossetal12,Kugland12}, and the 
question of whether shock acceleration, or the formation of non-thermal particle
populations occurs, is of fundamental importance. 

We have reviewed the necessary conditions that must be satisfied in 
order to achieve a clear detection of accelerated particles. Our analysis
confirms that several laser facilities currently operating may be capable 
of producing a shock which accelerates electrons to maximum energies where 
they can be clearly distinguished as shock accelerated particles. It is 
however quite unlikely that the same can be done for protons for any
currently existing facilities. We also note that
the requirement that the shock be completely collisionless can be relaxed
slightly, although it remains crucial that the 
magnetic Reynolds number is sufficiently large that the magnetic dissipation
can be neglected on the gyro-scale of the accelerating electrons.

There are a number of uncertainties in the analysis presented in this 
paper, about which we have tried to be transparent. However, 
we can summarise them here again. Firstly the question of injection, which is 
also of great importance in astrophysics. We have made the, not unreasonable, 
approximation that a small fraction of the shocked electrons 
are heated to a temperature 
on the order of $10\%$ that of the shocked ions. While this assumption is
of course arbitrary, if no acceleration is observed, it is still possible to inject 
fast electrons that satisfy the necessary conditions. On the other hand, should 
acceleration be detected without external injection, these experiments provide 
a novel platform to study the injection itself.

The other major uncertainty 
is the nature of the scattering $\omega_{\rm g} \tau_B$. This is an important topic 
in its own right in the study of collisionless shocks, and recent experiments 
are advancing our understanding \cite{Rossetal12,Kugland12}. However, as noted earlier,
the ratio of the gyroradius of a keV electron to the collisionless skin-depth 
is $r_{\rm g}({\rm keV})/(c/\omega_{\rm pe}) = 1.4 B_4^{-1} n_{16}^{1/2}$,
ie. on the correct scale to scatter resonantly the electrons. Future magnetised 
experiments will provide valuable information.

While the discussion in this paper has involved, in all cases, the 
presence of a strong large-scale field, 
there are ongoing efforts to realise collisionless 
unmagnetised shocks in the laboratory \cite{parketal12, Kugland12}.
Numerical simulations of unmagnetised shocks, in an astrophysical context, have 
developed rapidly in recent years \cite{kato08,spitkovsky08,martins09}, and are of 
vital importance to understanding the underlying kinetic processes. However, 
for the non-relativistic flow speeds we expect in the 
laboratory, our analysis suggests that a large scale magnetic field is required
to accelerate particles to energies where a detection can be made on the relevant
time-scale. The production of 
mildly-relativistic unmagnetised shocks in the laboratory has recently been 
demonstrated numerically 
\cite{fiuza}, using particle in cell simulations of intense 
($10^{20}-10^{22}{\rm W~cm}^{-2}$) laser pulses in an over-dense plasma. 
This is an exciting line of research with many applications,
however, regarding shock acceleration, the finite life-time and more importantly
the finite transverse extent of the shock front will be a limiting 
factor when distinguishing accelerated particles from the relativistic 
shock heated particles.

In conclusion, it appears quite possible that diffusive shock 
acceleration can be reproduced in the laboratory. 
Even with the help of a mega-Joule laser such as NIF, an unambiguous
detection of shock accelerated electrons will not be trivial, and is unlikely 
to be found without careful diagnostics and analysis, but should indeed be 
possible in the very near future. The success of such an experiment would 
be a first step in helping bring new insight to how Nature accelerates the 
most energetic particles in the universe.
  
\ack 

\vspace{-0.5cm}
{\footnotesize The research leading to these results has received funding from the 
European Research Council under the 
European Community's Seventh Framework Programme (FP7/2007-2013) / ERC grant 
agreement no. 247039 and no. 256973. BR gratefully acknowledges discussion 
with G Giacinti and C Ridgers.}

\end{document}